# Asymmetric Negative Refraction in Nonlocal Double-Wire Metamaterials

Tiago A. Morgado[1*], Sylvain Lannebère[1], Enrico M. Renzi[2,3], Andrea Alù[2,3], and Mário G. Silveirinha[4]

[1]*Instituto de Telecomunicações and Department of Electrical and Computer Engineering, University of Coimbra, 3030-290 Coimbra, Portugal*

[2]*Photonics Initiative, Advanced Science Research Center, City University of New York, New York, New York 10031, USA*

[3]*Physics Program, The Graduate Center, City University of New York, New York, New York 10026, USA*

[4]*University of Lisbon–Instituto Superior Técnico and Instituto de Telecomunicações, Avenida Rovisco Pais, 1, 1049-001 Lisboa, Portugal*

*E-mails:* tiago.morgado@co.it.pt, mario.silveirinha@tecnico.ulisboa.pt

## Abstract

We demonstrate that heterogeneous double-wire metamaterials—comprising two dissimilar and nonconnected wire arrays—enable loss-asymmetric hyperbolic dispersion and strongly asymmetric negative refraction. We unveil that, owing to their strongly nonlocal response and distinctive microstructure, such geometry supports two independent hyperbolic propagation channels with unequal losses, which can be selectively excited by free-space propagating waves incident at opposite angles. This angular selectivity gives rise to asymmetric negative refraction at the air-metamaterial interfaces, characterized by strong angular asymmetry in transmission and absorption, while preserving reciprocity. This phenomenon is scalable and can be realized with metallic wire arrays over a broad frequency spectrum extending from microwave to infrared frequencies, showing that nonlocality can emulate shear-like dissipative asymmetry that, in local media, requires lower spatial symmetry. Our findings open new avenues for directional energy transport and angle-selective wave control in photonic platforms.

[*] To whom correspondence should be addressed: E-mail: tiago.morgado@co.it.pt

Controlling electromagnetic wave propagation is central to modern photonics [1-3]. Metamaterials [4] and metasurfaces [5-6] provide a powerful route to precise control over light. Among them, hyperbolic metamaterials [7-10] and metasurfaces [10-11], also known as indefinite anisotropic media [12], form a particularly important class: their open isofrequency contours (IFCs), arising from opposite signs of the principal permittivity and/or permeability components, support a continuum of high-$k$ propagating modes and a diverging density of photonic states over broad bandwidths [7, 13-14]. This response underlies phenomena such as Purcell enhancement [7, 15-16], negative refraction [17-19], hyperlensing [20-24] and super-Planckian thermal emission [25-26].

While conventional hyperbolic media are uniaxial and mirror-symmetric, recent work has explored low-symmetry systems supporting hyperbolic shear responses, including low-symmetry crystals [27-30] and hyperbolic shear metasurfaces [31-33]. In these platforms, non-orthogonal and detuned resonances give rise to intrinsically non-diagonalizable dielectric tensors with imaginary off-diagonal (shear) components that cannot be removed by any coordinate transformation. As a result, these systems support strongly confined hyperbolic shear polaritons with tilted IFCs, frequency-dependent optical-axis rotation (axial dispersion), and pronounced asymmetric dissipation.

This behavior can be understood in terms of the distinct roles played by the dispersive and dissipative components of the dielectric response. For a reciprocal medium, the permittivity tensor can be written as $\boldsymbol{\varepsilon} = \boldsymbol{\varepsilon}' + i\boldsymbol{\varepsilon}''$, where $\boldsymbol{\varepsilon}' = \frac{\boldsymbol{\varepsilon} + \boldsymbol{\varepsilon}^\dagger}{2}$ is the Hermitian part that governs wave dispersion, and $\boldsymbol{\varepsilon}'' = \frac{\boldsymbol{\varepsilon} - \boldsymbol{\varepsilon}^\dagger}{2i}$ describes material losses. In local media, $\boldsymbol{\varepsilon}'$ determines the IFCs, while $\boldsymbol{\varepsilon}''$ controls the damping of the corresponding modes. Since the local dispersion relation is determined by a quadratic form, the IFCs in a plane are conic sections, such as ellipses, circles, or hyperbolas. Therefore, regardless of the specific structure of $\boldsymbol{\varepsilon}'$, these contours always possess two symmetry axes, defined by the directions that diagonalize $\boldsymbol{\varepsilon}'$. Although these axes may rotate with frequency, they remain well defined. The hyperbolic shear response first explored in [27] relies on the fact that, in low-symmetry systems, the axes associated with $\boldsymbol{\varepsilon}'$ and $\boldsymbol{\varepsilon}''$ do not coincide. This misalignment preserves the symmetry of the IFCs, and therefore of wave propagation, while inducing directional asymmetry in the losses, resulting in unequal attenuation on opposite sides of the same hyperbolic branch.

Beyond these resonance-driven indefinite anisotropic media, hyperbolic dispersion can also arise through nonresonant mechanisms. A prominent example is the double-wire metamaterial formed by two nonconnected crossed arrays of metallic wires, where hyperbolic dispersion originates from the microstructured geometry and the strongly nonlocal (spatially dispersive) response of the system [34-37]. Crucially, this geometry and the associated strong nonlocal response lead to a distinct topology in the IFCs: unlike the single hyperbola characteristic of conventional hyperbolic media [12], the double-wire metamaterial features contours consisting of two separate hyperbolas [35-36]. Remarkably, this nonresonant hyperbolic regime enables all-angle broadband negative refraction and partial focusing of electromagnetic waves using a planar slab [35-37].

In this Letter, we introduce double-wire metamaterials, formed by two dissimilar and nonconnected wire arrays, as a reciprocal platform that enables loss-asymmetric hyperbolic dispersion and asymmetric negative refraction. In contrast to shear-type hyperbolic systems, the asymmetry does not rely on frequency-dependent rotations of the principal axes or on low structural symmetry, but instead arises from the combination of strong nonlocality and material asymmetry in the crossed-wire geometry. The asymptotes of the hyperbolic branches are fixed by the wire geometry and do not rotate with frequency, so that the response exhibits no axial dispersion. Notably, this loss asymmetry can occur even when the dispersive and dissipative responses share the same symmetry axes, demonstrating that nonlocality can emulate shear-like dissipative responses without requiring low structural symmetry. We show that pairing a highly conducting wire array with a strongly lossy one creates two distinct propagation channels with markedly different attenuation, which can be selectively excited by waves incident at opposite angles. This mechanism produces angularly selective transmission and absorption, yielding strongly asymmetric negative refraction over a broad spectral range.

We consider heterogeneous double-wire metamaterials composed of two nonconnected arrays of metallic wires, oriented along the directions defined by the unit vectors $\hat{\mathbf{u}}_1 = \sin\theta_1 \hat{\mathbf{u}}_x + \cos\theta_1 \hat{\mathbf{u}}_z$ and $\hat{\mathbf{u}}_2 = -\sin\theta_2 \hat{\mathbf{u}}_x + \cos\theta_2 \hat{\mathbf{u}}_z$ (Fig. 1). The two wire arrays may differ in wire radii ($r_{\mathrm{w},1} \neq r_{\mathrm{w},2}$) and/or in the constituent metallic material ($\varepsilon_{\mathrm{m},1} \neq \varepsilon_{\mathrm{m},2}$). The symbols $\varepsilon_{\mathrm{m},1}$ and $\varepsilon_{\mathrm{m},2}$ denote the complex relative permittivities of the metallic wires in arrays 1 and 2, respectively. These permittivities are described by Drude dispersion models, $\varepsilon_{\mathrm{m},l} = \varepsilon_{\infty,l} - \omega_{\mathrm{p},l}^2 \big/ \left(\omega^2 + i\omega\Gamma_l\right)$, with $l = 1,2$. The wire arrays are embedded in a host medium with relative permittivity $\varepsilon_{\mathrm{h}}$, taken as unity throughout this work for simplicity.

In the long-wavelength regime, the electromagnetic response of the metamaterial is characterized by the nonlocal permittivity $\overline{\overline{\varepsilon}}_{\mathrm{eff}} = \varepsilon_{\mathrm{h}} \left[ \overline{\overline{\mathbf{I}}} + \left(\varepsilon_{11} - 1\right) \hat{\mathbf{u}}_1 \hat{\mathbf{u}}_1 + \left(\varepsilon_{22} - 1\right) \hat{\mathbf{u}}_2 \hat{\mathbf{u}}_2 \right]$ [37, 38-41], where $\overline{\overline{\mathbf{I}}}$ is the identity dyadic. The components $\varepsilon_{11}$ and $\varepsilon_{22}$ are given by [41-42]

$$\varepsilon_{ll}\left(\omega,k_l\right)=1+\frac{1}{\dfrac{1}{(\varepsilon_{\mathrm{m},l}-1)f_{\mathrm{V},l}}-\dfrac{\varepsilon_{\mathrm{h}}k_0^2-k_l^2}{\beta_{\mathrm{p},l}^2}},\quad l=1,2, \tag{1}$$

where $\omega$ is the angular frequency, $c$ is the speed of light in vacuum, $k_0=\omega/c$ is the free-space wavenumber, $f_{\mathrm{V},l}=\pi(r_{\mathrm{w},l}/a)^2$ is the metal volume fraction, $r_{\mathrm{w},l}$ the wire radius, $a$ the lattice constant, $\beta_{\mathrm{p},l}=[2\pi/(\ln(a/2\pi r_{\mathrm{w},l})+0.5275)]^{1/2}/a$ the plasma wavenumber, and $k_l=\mathbf{k}\cdot\hat{\mathbf{u}}_l$ ($l=1,2$) with $\mathbf{k}=\left(k_x,k_y,k_z\right)$. The index $l$=1,2 identifies the wire subarray. The wave-vector dependence in Eq. (1) reflects the nonlocal nature of wire media. Although the lattice constant $a$ is subwavelength and allows a homogenized description, the wire length is not required to be subwavelength. As a result, electric currents can extend along the wires over distances larger than $a$, giving rise to strong spatial dispersion [38-41].

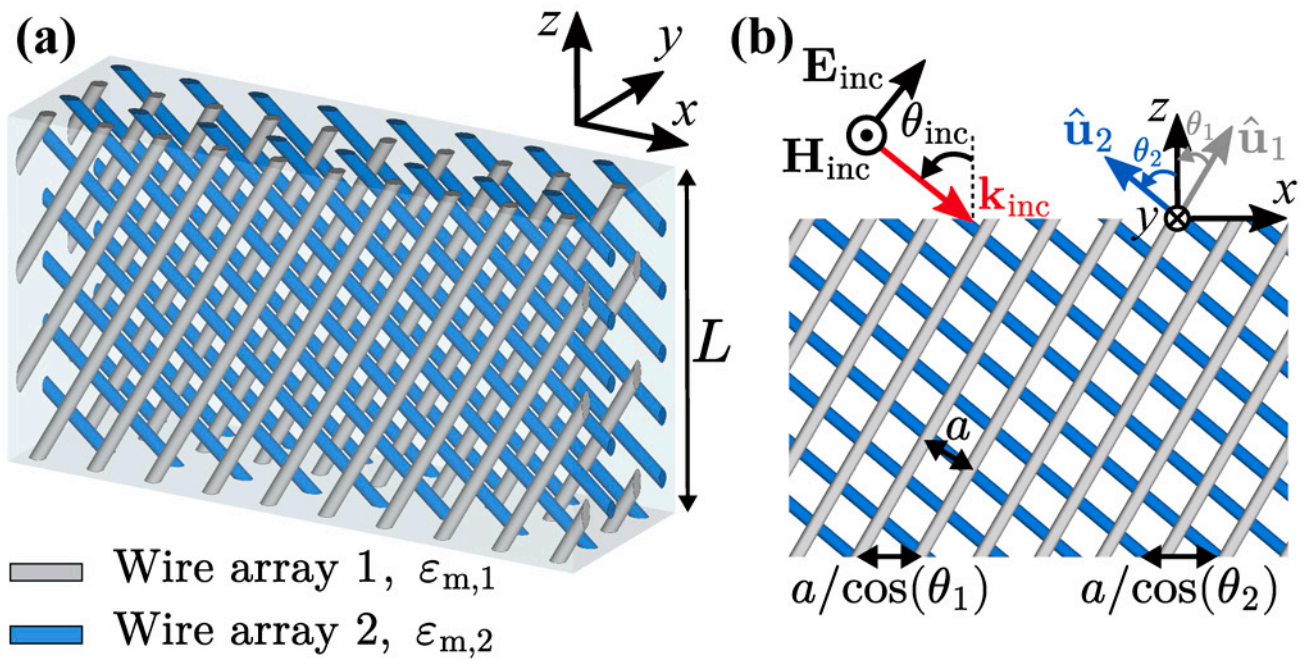

**Figure 1.** Geometry of the heterogeneous double-wire medium composed of two distinct and nonconnected arrays of metallic wires (wire array 1 and 2). Array 1 comprises wires oriented along $\hat{\mathbf{u}}_1=\sin\theta_1\hat{\mathbf{u}}_x+\cos\theta_1\hat{\mathbf{u}}_z$, and array 2 along $\hat{\mathbf{u}}_2=-\sin\theta_2\hat{\mathbf{u}}_x+\cos\theta_2\hat{\mathbf{u}}_z$. The wires in array 1 and array 2 may have different radii ($r_{\mathrm{w},1}\neq r_{\mathrm{w},2}$) and/or be made of different materials ($\varepsilon_{\mathrm{m},1}\neq\varepsilon_{\mathrm{m},2}$). The distance between each adjacent set of wires is $a/2$. The plane of incidence is the *xz* plane and the incident wave is TM-*z* polarized. (a) Perspective view; (b) Front view (*xz* plane).

We focus on wave propagation in the *xz* plane ($k_y=0$), with magnetic field polarized along the *y*-direction (transverse magnetic (TM) polarization) [Fig. 1(b)]. As shown in Ref. [43], the dispersion relation for TM plane-waves supported by the double-wire medium is a sixth-degree polynomial equation in the variable $k_z$, due to the underlying nonlocality. Thus, the bulk metamaterial supports six independent plane-wave modes with $y$-polarized magnetic field, corresponding to three pairs of counter-propagating waves. The existence of multiple modes at the same frequency is a direct manifestation of the nonlocal response of the metamaterial, since in local media each polarization is associated with only one pair of counterpropagating plane waves.

Importantly, in the low-frequency regime and for weak dissipation, only two plane-wave modes can propagate, while the remaining four modes are evanescent. In Fig. 2, we show the IFCs of the propagating modes for $\theta_1 = 0^\circ$ and varying $\theta_2$.

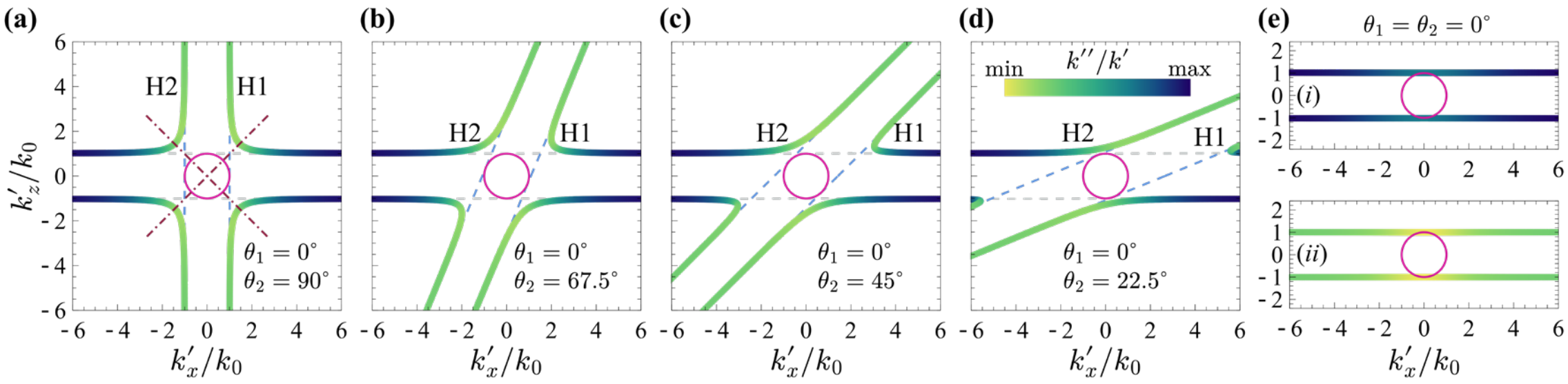

**Figure 2.** IFCs of the propagating plane-wave modes supported by the heterogeneous double-wire medium at $\omega/(2\pi) = 20$ THz, with $\theta_1 = 0^\circ$ and $\theta_2$ varying from $90^\circ$ (orthogonal wire arrays) to $0^\circ$ (parallel wire arrays). (a) $\theta_2 = 90^\circ$; (b) $\theta_2 = 67.5^\circ$; (c) $\theta_2 = 45^\circ$; (d) $\theta_2 = 22.5^\circ$; (e) $\theta_2 = 0^\circ$, with (*i*) the mode associated with wire array 1 and (*ii*) the mode associated with wire array 2. Magenta circles denote the free-space IFCs. Gray and blue dashed lines represent the asymptotes of the two hyperbolas, whereas the dark-red dash-dotted lines in (a) mark the asymptote bisectors. The two hyperbolas are labeled H1 and H2. The lattice constant is $a = 2.5\ \mu$m and the wire radii are $r_{\mathrm{w},1} = r_{\mathrm{w},2} = 0.05a$. The wires of the arrays 1 and 2 follow a Drude model with plasma frequencies $\omega_{\mathrm{p},1}/(2\pi) = 2000$ THz and $\omega_{\mathrm{p},2} = 2\omega_{\mathrm{p},1}$, respectively, equal damping rates $\Gamma_1 = \Gamma_2 = 0.02\omega_{\mathrm{p},1}$, and high-frequency permittivities $\varepsilon_{\infty,1} = \varepsilon_{\infty,2} = 1$.

Consistent with previous works [34-37], Figs. 2(a)-(d) show that for configurations with non-parallel wire arrays, the double-wire medium supports propagating modes with hyperbolic IFCs. In contrast to conventional hyperbolic media, whose IFCs consist of a single hyperbola [12], in this nonlocal wire metamaterial the IFCs consist of two distinct hyperbolas, with one pair of asymptotes aligned with the direction perpendicular to the wire array 1 [see gray dashed lines in Figs. 2(a)-(d)] and the other aligned with the direction perpendicular to the wire array 2 [see blue dashed lines in Figs. 2(a)-(d)], i.e., directions perpendicular to $\hat{\mathbf{u}}_1$ and $\hat{\mathbf{u}}_2$. This is a key consequence of nonlocality: the IFCs are no longer restricted to the conic sections allowed in local media, such as a single ellipse or hyperbola. Instead, the spatially dispersive response enables a richer topology that cannot be reproduced by a local permittivity tensor.

The two asymptotes coincide with the flat IFCs of the quasi-TEM modes supported by each isolated wire array [44]. Physically, near these asymptotes the wave vector is large and the two wire arrays become weakly coupled, so that each hyperbolic arm approaches the dispersion of the corresponding isolated array. As $\theta_2$ varies from $90^\circ$ to $0^\circ$, the asymptote associated with array 2 rotates accordingly, which changes the aperture angle of the hyperbolas and results in a progressive rotation of the IFCs. Even when the two wire arrays are made of different metals, as in Fig. 2, the bisectors of the asymptote directions remain approximate symmetry axes of the IFCs, although they are not exact symmetry axes of the macroscopic permittivity. In the limiting case of parallel wire arrays [$\theta_1 = \theta_2 = 0^\circ$, Fig. 2(e)], the two hyperbolas collapse into a pair of flat contours corresponding to two TEM modes, each associated with one of the parallel wire arrays.

Notably, the heterogeneity between the two wire arrays leaves the dispersion contours approximately unaltered, as discussed above, but breaks the symmetry of the loss distribution. Because the two arrays are distinct, the different arms of the hyperbolas are linked to distinct loss channels, so that opposite arms exhibit unequal attenuation [see Figs. 2(a)-(d)]. This gives rise to hyperbolic-like IFCs with nearly symmetric dispersion but asymmetric dissipation [Figs. 2(a)-(d)], reminiscent of hyperbolic shear dispersions. The dependence of this loss redistribution on the angle between the two wire arrays is analyzed further in Ref. [43].

Even though the loss-asymmetric hyperbolic dispersion contours in Figs. 2(a)-(d) resemble the hyperbolic contours with asymmetric dissipation reported for low-symmetry crystals [27-30] and for hyperbolic shear metasurfaces [31-32], their origin is fundamentally different. In those local indefinite media, asymmetric dissipation is associated with non-orthogonal detuned and non-simultaneously diagonalizable dispersive and dissipative responses [27, 32-33]. By contrast, in the heterogeneous double-wire medium, it originates from the strongly nonlocal response of the metamaterial, which provides two propagation channels with unequal loss.

A direct consequence of these distinct physical origins is that asymmetric dissipation emerges in the heterogeneous double-wire medium even for orthogonal wire arrays, as shown in Fig. 2(a). In this case, the structure has two exact symmetry axes in the propagation plane, aligned with the wire directions $\hat{\mathbf{u}}_1$ and $\hat{\mathbf{u}}_2$, corresponding to the $x$ and $z$ axes in Fig. 2(a). Accordingly, the IFCs exhibit the same parity symmetry, both in their dispersion and dissipation. However, this symmetry does not relate to opposite arms of the same hyperbola; instead, it maps one hyperbola onto the other. Thus, each individual hyperbola can still display unequal attenuation between its arms, even though the full set of contours respects the symmetry of the structure.

This behavior cannot emerge in low-symmetry crystals [27-30] or hyperbolic shear metasurfaces [31-33] under orthogonal configurations, for which the shear is zero. In this limit, the material response becomes fully diagonalizable, so that shear anisotropy and off-diagonal loss terms vanish, yielding hyperbolic contours with symmetric dissipation. By contrast, in the present metamaterial, nonlocality introduces additional degrees of freedom: the exact symmetry axes constrain the pair of hyperbolas as a whole, without forcing each hyperbola to be loss-symmetric. The approximate symmetry axes, defined by the bisectors of the asymptotes of the two hyperbolic branches (see dot-dashed lines in Fig. 2(a)) are also preserved in the dispersion contours, while the loss imbalance between the arms of each hyperbola remains (Fig. 2(a)). This shows that a system with macroscopic inversion and mirror symmetries can reproduce a shear-like asymmetric dissipative response, provided that the response is nonlocal.

Another distinctive feature of the proposed heterogeneous double-wire medium is that its propagation directions are set by geometry alone. Unlike low-symmetry crystals [27-30] and hyperbolic shear metasurfaces [31-33], it exhibits no axial dispersion. The asymptotes of the hyperbolic branches are fixed by the wire geometry, remaining perpendicular to the wire directions and coinciding with the frequency-independent flat TEM IFCs supported by each isolated wire array [44-45]. Consequently, the propagation directions are selected by design and remain invariant across the entire frequency band.

Unlike conventional shear hyperbolic media [27-32-33], where asymmetric dissipation requires highly confined guided modes, our nonlocal metamaterial exhibits pronounced attenuation asymmetry already at small transverse wave vectors [Figs. 2(a)-(e)]. The strongly nonlocal response extends hyperbolic dispersion to the low-*k* region. This opens a direct route to coupling free-space radiation to asymmetrically channeled waves. This is exemplified in Fig. 3(a), which shows the hyperbolic IFCs of the propagating modes in a heterogeneous double-wire medium with $\theta_1 = \theta_2 = 45^\circ$ (orthogonal arrays), along with the free-space IFCs (magenta circle), for $\omega/(2\pi) = 20$ THz. Here, we consider a potentially realizable configuration operating at infrared (IR) frequencies, formed by two Ag wire arrays with unequal radii. Both Ag wire arrays are described by the same Drude model with $\varepsilon_{\infty,l} = 5$, $\omega_{\mathrm{p},l}/(2\pi) = 2175$ THz, and $\Gamma_l = 0.002\omega_{\mathrm{p},l}$ [46-48], for $l = 1,2$. This design exploits the fact that Ag wires behave as good conductors when their radius is much larger than the metal skin depth ($r_w >> \delta_{Ag}$), whereas they become highly lossy when $r_w < \delta_{Ag}$. Accordingly, one array is formed by thick (low-loss) wires ($r_{w,1} > \delta_{Ag}$), while the other is composed of thin (highly lossy) wires ($r_{w,2} < \delta_{Ag}$).

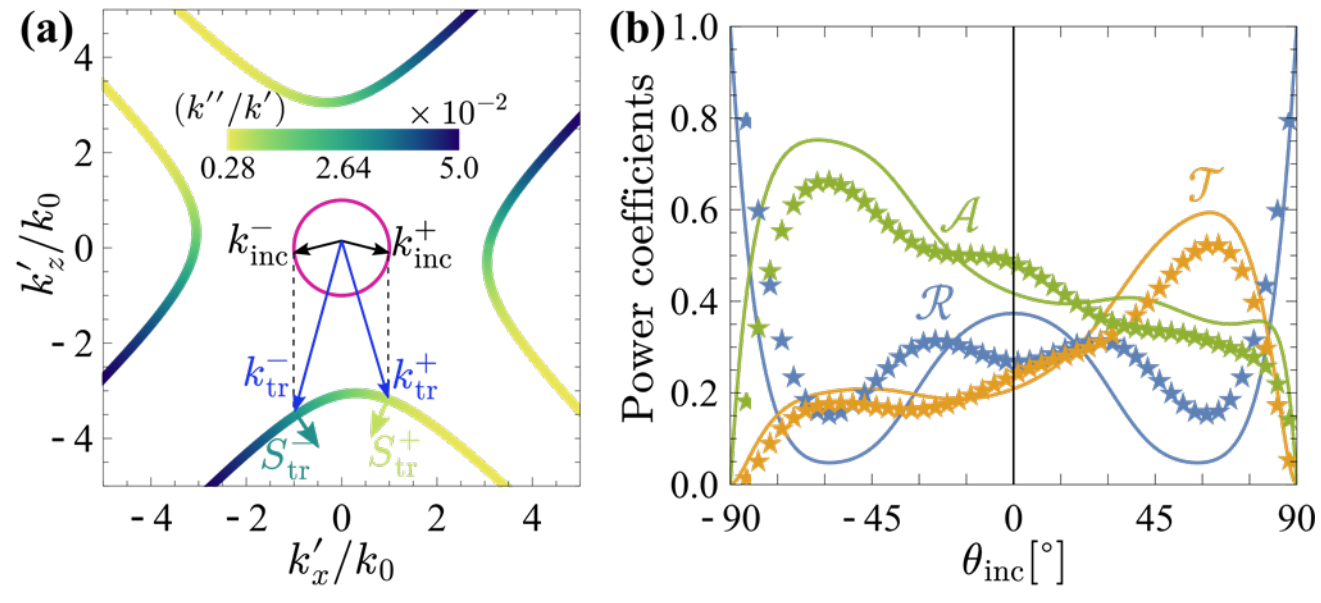

**Figure 3.** Dispersion, dissipation, and plane-wave response of an Ag double-wire medium in the IR regime. The system consists of two distinct Ag wire arrays: array 1 with $r_{w,1} = 150$ nm and array 2 with $r_{w,2} = 15$ nm. The lattice constant is $a = 1\ \mu$m. (a) IFCs of the fundamental plane wave mode supported by the metamaterial (yellow-blue-green gradient), as well as the free-space IFC (magenta color), at $\omega/(2\pi) = 20$ THz and with $\theta_1 = \theta_2 = 45°$. The transmitted wave vectors $\mathbf{k}_{tr}^{\pm}$ are determined by the conservation of the tangential component of the wave vector $\mathbf{k}$, whereas the Poynting vectors $\mathbf{S}_{tr}^{\pm}$ are normal to the IFC and point toward increasing frequency. (b) Reflectance ($\mathcal{R}$), transmittance ($\mathcal{T}$), and absorptance ($\mathcal{A}$) versus the angle of incidence ($\theta_{inc}$). The thickness of the slab is $L = 20a$, and the remaining parameters are as in (a). Solid lines: nonlocal homogenization results; discrete symbols (stars): full-wave results [49].

In this IR implementation, the large contrast in wire radii causes the hyperbolic IFCs to deviate noticeably from exact mirror symmetry about the *x*- and *z*-axes. Nevertheless, the underlying mechanism remains unchanged: the two hyperbolic propagation channels persist, while the geometric contrast produces a marked imbalance in their effective attenuation.

Remarkably, the proposed metastructure enables asymmetric negative refraction for all incidence angles, as shown in Fig. 3(a). Crucially, the wave attenuation is markedly asymmetric: waves with $k_x' > 0$ are only weakly damped, whereas those with $k_x' < 0$ exhibit significantly larger attenuation, revealing a clear directional imbalance in losses. Related shear-induced asymmetric negative refraction has been discussed in Ref. [31].

The asymmetric scattering is quantified in Fig. 3(b) with reflectance ($\mathcal{R}$), transmittance ($\mathcal{T}$), and absorptance ($\mathcal{A}$). The quantities $\mathcal{R}$, $\mathcal{T}$, and $\mathcal{A}$ are shown as functions of the angle of incidence ($\theta_{\mathrm{inc}}$), calculated using both our nonlocal homogenization model (solid curves; details on the calculation can be found in Ref. [43]) and full-wave simulations performed in CST Studio Suite [49] (discrete symbols). The solid curves and symbols show good overall qualitative agreement, particularly in capturing the angular asymmetry of transmission and absorption. The slight deviations reflect the limits of homogenization when the wire radius becomes smaller than the metal skin depth.

Notably, Fig. 3(b) shows that the considered geometry exhibits a clear angular asymmetry in transmission and absorption, while the reflectance remains symmetric, as required by reciprocity. Waves with $\theta_{\mathrm{inc}} < 0$ experience enhanced absorption, while those with $\theta_{\mathrm{inc}} > 0$ are transmitted more effectively. Thus, the structure exhibits angle-selective transmission and absorption at IR frequencies. The reciprocity constraints on the scattering parameters are discussed in Ref. [43].

Importantly, increasing the angle between the two wire arrays ($\theta_0 = \theta_1 + \theta_2$) enables free-space propagating waves to couple to hyperbolic modes with larger attenuation contrast [see Fig. 4(a) and Fig. 3(a)]. Consequently, for $\theta_0 > 90^{\circ}$, the angular asymmetry in transmission and absorption is strongly enhanced, as shown in Fig. 4(b). Waves with $\theta_{\mathrm{inc}} < 0$ are strongly absorbed, with vanishingly small transmittance, whereas those with $\theta_{\mathrm{inc}} > 0$ are effectively transmitted. Remarkably, this enhanced asymmetry is achieved even with considerably thinner metamaterial slabs [see Figs. 3(b) and 4(b)]. This occurs because, as $\theta_0$ increases, the wires become more aligned with the interfaces, allowing waves to traverse longer effective electrical paths within the lossy array even for thinner slabs. As a result, the structure operates as an angularly selective absorber–transmitter at IR frequencies, enabling strongly asymmetric control of energy flow.

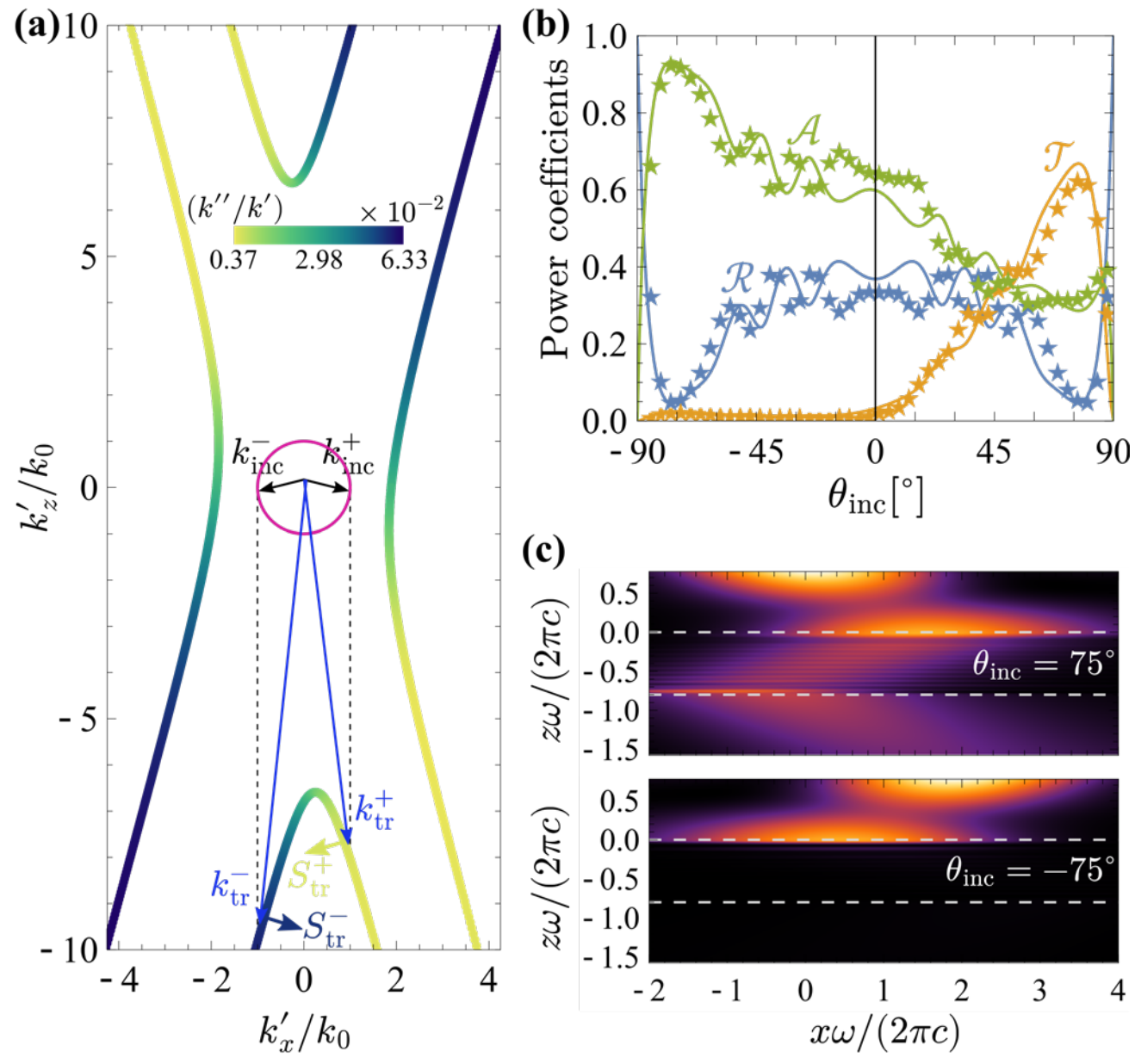

**Figure 4.** Enhanced asymmetric response of the Ag double-wire medium in the IR regime. The wire radii, lattice constant, and frequency are as in Fig. 3; here $\theta_1 = \theta_2 = 75°$. The legend of (a) and (b) is as in Fig. 3. (a) IFC and dissipation map. (b) $\mathcal{R}$, $\mathcal{T}$, and $\mathcal{A}$ versus $\theta_{\rm inc}$ for $L = 12a$. (c) Normalized magnetic-field intensity ($|\mathbf{H}|^2$) for an incident TM-polarized Gaussian beam with beam waist $w_0 = 4\pi c/\omega$, calculated using the nonlocal effective medium model. The metamaterial slab has the same geometry as in (b). The dashed lines represent the air interfaces.

Figure 4(c) show Gaussian-beam excitations at representative positive and negative incidence angles, with calculation details and additional results provided in Ref. [43]. For $\theta_{\rm inc} > 0$, the beam undergoes negative refraction at both interfaces and is efficiently transmitted through the metamaterial slab, as it primarily interacts with the low-loss propagation channel associated with array 1, formed by thick Ag wires. Conversely, for $\theta_{\rm inc} < 0$ the beam couples mainly to the highly dissipative channel associated with array 2, composed of thin Ag wires, and is therefore strongly attenuated inside the slab. These results clearly demonstrate asymmetric negative refraction in the heterogeneous double-wire medium at IR frequencies.

In the End Matter, we further show, through a representative Ag/InSb implementation, that the loss-asymmetric mechanism and asymmetric negative refraction can be extended to microwave frequencies.

In summary, we have demonstrated that heterogeneous double-wire metamaterials provide a versatile platform for achieving loss-asymmetric hyperbolic dispersion and asymmetric negative refraction across both microwave and IR frequencies. Using nonlocal homogenization theory and full-wave simulations, we have shown that pairing two dissimilar wire arrays, one highly conducting and the other more lossy, creates two independent propagation channels with unequal attenuation, which can be selectively excited by free-space propagating waves incident at opposite angles. Increasing the angle between the wire arrays enhances the attenuation contrast across the two propagation channels, leading to a more pronounced angular asymmetry in transmission and absorption. Importantly, this response is compatible with high-symmetry configurations, including macroscopic inversion and parity symmetries, because nonlocality provides additional degrees of freedom beyond those available in local media. In this sense, heterogeneous double-wire metamaterials can reproduce shear-like asymmetric dissipative effects that, in local systems, typically require low structural symmetry and strongly frequency-dependent optical axes. These findings reveal that engineered spatial dispersion and loss imbalance can give rise to highly asymmetric energy transport in fully reciprocal media, opening new avenues for angle-selective photonic devices and directional thermal emitters.

**Acknowledgments:** This work was funded by FCT – Fundação para a Ciência e a Tecnologia, I.P., under project AWMETA (reference: 2024.17646.PEX, DOI: 10.54499/2024.17646.PEX), and by the Simons Foundation (Award SFI-MPS-EWP-00008530-10). Further support was provided by FCT and, when eligible, by EU funds under project UID/50008/2025 (Instituto de Telecomunicações, DOI: 10.54499/UID/50008/2025).

## End Matter

The physical mechanism underlying loss-asymmetric hyperbolic dispersion and asymmetric negative refraction extends to the microwave regime, as we demonstrate here for a heterogeneous double-wire medium formed by two wire arrays made of different materials. Specifically, array 1 consists of Ag wires, whereas array 2 consists of InSb wires described by a Drude model with $\varepsilon_{\infty,2} = 15.7$, $\omega_{p,2}/(2\pi) = 7.959$ THz, and $\Gamma_2 = 0.033\omega_{p,2}$, consistent with room-temperature InSb [50].

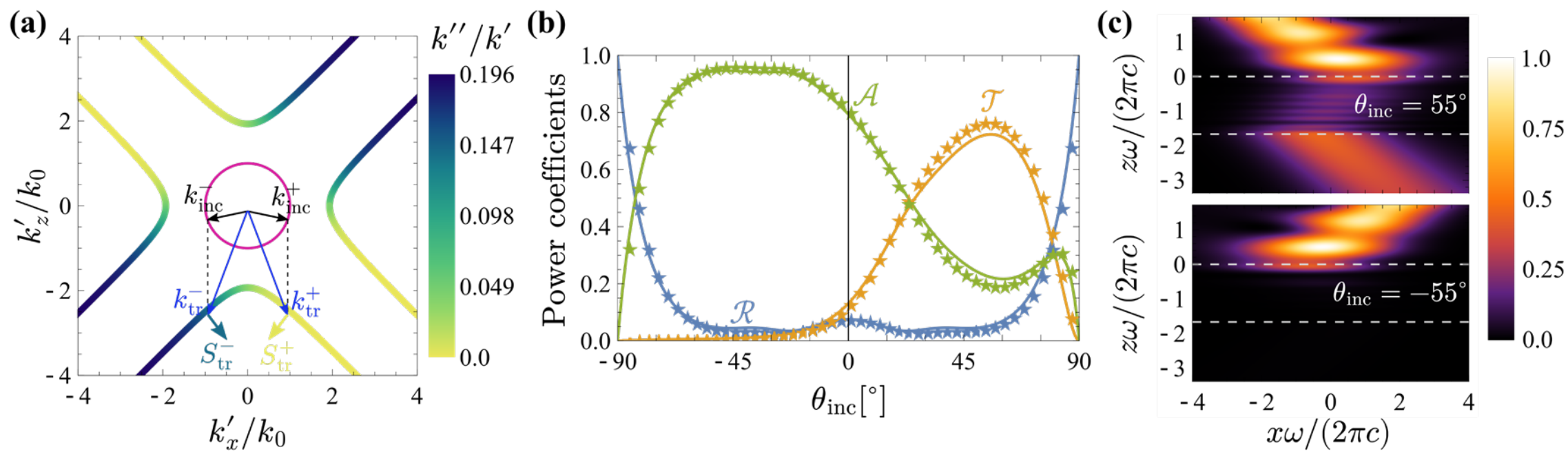

**Figure 5.** Dispersion, dissipation, and plane-wave/Gaussian-beam responses of an Ag/InSb double-wire medium in the microwave regime. Array 1 is formed by Ag wires with $r_{w,1} = 125\ \mu$m, and array 2 is formed by InSb wires with radius $r_{w,2} = 25\ \mu$m. The lattice constant is $a = 2.5$ mm. (a) IFCs of the fundamental plane wave mode supported by the Ag/InSb double-wire medium (yellow-blue-green gradient), as well as the free-space IFC (magenta color), at $\omega/(2\pi) = 20$ GHz and with $\theta_1 = \theta_2 = 45°$. (b) Reflectance ($\mathcal{R}$), transmittance ($\mathcal{T}$), and absorptance ($\mathcal{A}$) versus the angle of incidence ($\theta_{inc}$). The thickness of the slab is $L = 10a$, and the remaining parameters are as in (a). Solid lines: nonlocal homogenization results; discrete symbols (stars): full-wave results [49] (c) Normalized magnetic-field intensity ($|\mathbf{H}|^2$) inside and outside the Ag/InSb double-wire medium calculated using the nonlocal effective medium model. The excitation is a TM-polarized Gaussian beam with beam waist $w_0 = 4\pi c/\omega$. All other parameters are identical to those in panels (a) and (b). The metamaterial slab is periodic along the $x$- and $y$- directions. The dashed lines represent the interfaces of the slab.

Figures 5 and 6 show results for two representative microwave configurations. Figure 5 corresponds to $\theta_1 = \theta_2 = 45°$, $L = 10a$, and $f = 20$ GHz, whereas Fig. 6 corresponds to $\theta_1 = \theta_2 = 67.5°$, $L = 6a$, and $f = 13$ GHz. The left panels, Figs. 5(a) and 6(a), show the IFCs of the propagating hyperbolic modes. As in the IR configurations [see Figs. 3(a) and 4(a)], the IFCs of Figs. 5(a) and 6(a) show that all incoming waves couple to negatively refracted hyperbolic modes, independently of the incidence angle, while retaining a strongly asymmetric attenuation profile with respect to the sign of $k'_x$. Waves with $k'_x > 0$ couple predominantly to the low-loss Ag-wires channel and are only weakly attenuated, whereas waves with $k'_x < 0$ couple mainly to the lossy InSb-wires channel and undergo strong attenuation. Thus, the microwave implementation reproduces the same loss-asymmetric hyperbolic dispersion mechanism observed at IR frequencies. Unlike the IR configurations of Figs. 3(a) and 4(a), however, the dispersion contours appear mirror-symmetric about the *x*- and *z*-axes, with any residual asymmetry being barely discernible.

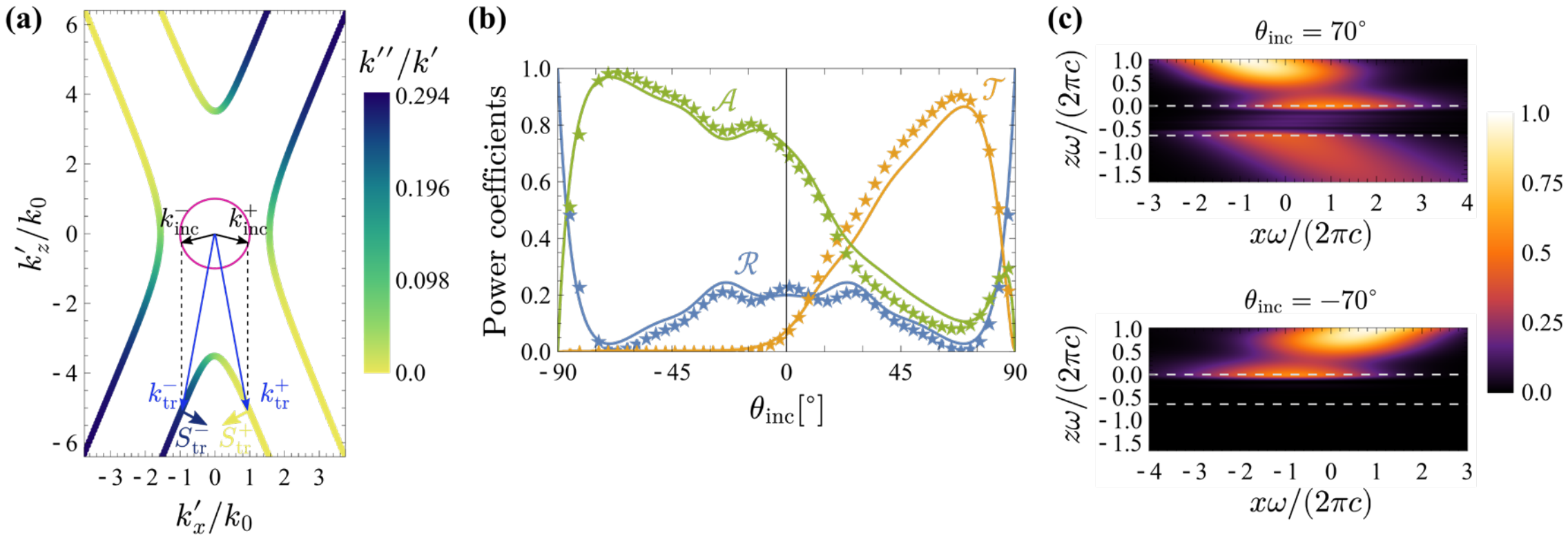

**Figure 6.** Similar to Fig. 5, except that $\omega/(2\pi) = 13$ GHz, $\theta_1 = \theta_2 = 67.5^\circ$, and $L = 6a$.

The central panels, Figs. 5(b) and 6(b), show the reflectance, transmittance, and absorptance as functions of the incidence angle, calculated with the nonlocal homogenization model (solid curves) and compared with full-wave simulations carried out in CST Studio Suite [49] (discrete symbols). The excellent agreement between the solid curves and discrete symbols confirms the accuracy of the nonlocal homogenization model for the microwave implementation.

Consistent with the IR results, Figs. 5(b) and 6(b) show that, in the microwave regime, the heterogeneous double-wire medium also functions as an angle-dependent absorbing–transmitting platform. Waves incident at positive angles ($\theta_{\rm inc} > 0$) are transmitted with high efficiency, whereas those impinging from negative angles ($\theta_{\text{inc}} < 0$) undergo strong absorption. This pronounced angular contrast in transmission and absorption directly stems from the loss-asymmetric hyperbolic dispersion of the medium.

As in the IR regime, we also analyzed the refraction of a Gaussian beam at the interfaces of the heterogeneous double-wire metamaterial slab in the microwave regime, using the nonlocal homogenization procedure described in [43]. Figures 5(c) and 6(c) show the normalized squared magnetic-field amplitude for Gaussian beams incident at fixed positive and negative angles. The results demonstrate that the heterogeneous double-wire medium enables strongly asymmetric negative refraction at microwave frequencies: beams with positive angles of incidence ($\theta_{\text{inc}} > 0$) are negatively refracted at both interfaces and efficiently transmitted through the slab, whereas those arriving from negative angles ($\theta_{\text{inc}} < 0$) are strongly absorbed inside the slab. Furthermore, in Sec. F of Ref. [43], additional full-wave simulations show that this effect persists in finite-width slabs comprising only a few wire planes along *y*, highlighting the potential for compact and practically implementable double-wire metastructures.

Moreover, comparing Figs. 5(a) and 6(a) clearly shows that increasing the angle between the two wire arrays ($\theta_0 = \theta_1 + \theta_2$) enhances the attenuation contrast of the hyperbolic modes excited by free-space waves. This increased attenuation contrast translates into a stronger angular asymmetry in transmission and absorption for the $\theta_0 = 135^\circ$ configuration, even though the slab is considerably thinner ($L = 6a$) than in the configuration with $\theta_0 = 90^\circ$ ($L = 10a$) [see Figs. 5(b) and 6(b)]. As a result, configurations with larger $\theta_0$ enable pronounced asymmetric negative refraction even with thinner metamaterial slabs [see Figs. 5(c) and 6(c)]. This behavior is consistent with the IR results and arises from the longer effective electrical path traveled by the excited modes within the wire arrays as $\theta_0$ increases.

# Supplemental Material for

# "Asymmetric Negative Refraction in Nonlocal Double-Wire Metamaterials"

Tiago A. Morgado[1*], Sylvain Lannebère[1], Enrico M. Renzi[2,3], Andrea Alù[2,3], and Mário G. Silveirinha[4]

*[1]Instituto de Telecomunicações and Department of Electrical and Computer Engineering, University of Coimbra, 3030-290 Coimbra, Portugal*

*[2]Photonics Initiative, Advanced Science Research Center, City University of New York, New York, New York 10031, USA*

*[3]Physics Program, The Graduate Center, City University of New York, New York, New York 10026, USA*

*[4]University of Lisbon–Instituto Superior Técnico and Instituto de Telecomunicações, Avenida Rovisco Pais, 1, 1049-001 Lisboa, Portugal*

*E-mails:* tiago.morgado@co.it.pt, mario.silveirinha@tecnico.ulisboa.pt

This Supplemental Material provides additional theoretical derivations and numerical results supporting the findings presented in the main text. It is organized as follows.

**CONTENTS**

[*] To whom correspondence should be addressed: E-mail: tiago.morgado@co.it.pt

## A. DISPERSION RELATION

Here, we derive the dispersion relation of the plane-wave modes supported by the bulk heterogeneous double-wire medium illustrated in Fig. 1 of the main text. In the long-wavelength limit, the electromagnetic response of this medium can be characterized by the nonlocal permittivity dyadic $\overline{\overline{\varepsilon}}_{\text{eff}} = \varepsilon_{\text{h}}\left[\overline{\overline{\mathbf{I}}} + \left(\varepsilon_{11}-1\right)\hat{\mathbf{u}}_1\hat{\mathbf{u}}_1 + \left(\varepsilon_{22}-1\right)\hat{\mathbf{u}}_2\hat{\mathbf{u}}_2\right]$ [S1], where $\overline{\overline{\mathbf{I}}}$ is the identity dyadic, and the permittivity components $\varepsilon_{11}$ and $\varepsilon_{22}$ are given by [S2-S3],

$$\varepsilon_{ll}\left(\omega,k_l\right) = 1 + \frac{1}{\dfrac{1}{(\varepsilon_{\text{m},l}-1)f_{\text{V},l}} - \dfrac{\varepsilon_{\text{h}}k_0^2 - k_l^2}{\beta_{\text{p},l}^2}}, \quad l = 1,2, \tag{S1}$$

where $\omega$ is the angular frequency, $c = 1/\sqrt{\varepsilon_0\mu_0}$ is the speed of light in vacuum, $k_0 = \omega/c$ is the free-space wavenumber, $f_{\text{V},l} = \pi(r_{\text{w},l}/a)^2$ is the metal volume fraction of array $l$ ($l$=1,2), $r_{\text{w},l}$ the wire radius of array $l$, $a$ the lattice constant, and $\beta_{\text{p},l} = [2\pi/(\ln(a/2\pi r_{\text{w},l}) + 0.5275)]^{1/2}/a$ the plasma wavenumber of array $l$, and $k_l = \mathbf{k}\cdot\hat{\mathbf{u}}_l$ ($l = 1,2$) with $\mathbf{k} = \left(k_x, k_y, k_z\right)$.

Assuming an $e^{-i\omega t}$ time dependence and in the absence of sources, Maxwell's equations can be written as:

$$\begin{cases} \nabla\times\mathbf{E} = -\mu_0\dfrac{\partial\mathbf{H}}{\partial t} = i\omega\mu_0\mathbf{H} \\ \nabla\times\mathbf{H} = \varepsilon_0\overline{\overline{\varepsilon}}_{\text{eff}}\cdot\dfrac{\partial\mathbf{E}}{\partial t} = -i\omega\varepsilon_0\overline{\overline{\varepsilon}}_{\text{eff}}\cdot\mathbf{E} \end{cases}. \tag{S2}$$

Combining both Eqs. (S2) we obtain:

$$\nabla\times\nabla\times\mathbf{E} = k_0^2\overline{\overline{\varepsilon}}_{\text{eff}}\cdot\mathbf{E}. \tag{S3}$$

Equation (S3) can be written in the form

$$\left[k^2\overline{\overline{\mathbf{I}}} - \mathbf{k}\mathbf{k} - k_0^2\overline{\overline{\varepsilon}}_{\text{eff}}\right]\cdot\mathbf{E} = 0 \tag{S4}$$

where $k^2 = k_x^2 + k_y^2 + k_z^2$ and $\mathbf{kk}$ is the dyadic (outer) product of two vectors. The nontrivial solutions $(\omega, \mathbf{k})$ of the homogeneous system of equations (S4) occur when

$$\det\left[k^2\overline{\mathbf{I}} - \mathbf{kk} - k_0^2\overline{\overline{\varepsilon}}_{\text{eff}}\right] = 0 \tag{S5}$$

For propagation in the *xz*-plane ($\partial/\partial y = 0$ or $k_y = 0$) with the magnetic field along the *y*-direction (TM polarization), the dispersion relation (S5) reduces, after algebraic manipulation, to the following polynomial equation:

$$\begin{aligned}
&k_z^2\left(1+(\varepsilon_{11}-1)\cos^2(\theta_1)+(\varepsilon_{22}-1)\cos^2(\theta_2)\right)+\\
&2k_xk_z\left((\varepsilon_{11}-1)\cos(\theta_1)\sin(\theta_1)-(\varepsilon_{22}-1)\cos(\theta_2)\sin(\theta_2)\right)+\\
&k_x^2\left(1+(\varepsilon_{11}-1)\sin^2(\theta_1)+(\varepsilon_{22}-1)\sin^2(\theta_2)\right)+\\
&\frac{1}{2}k_0^2\varepsilon_{\text{h}}\left(-(\varepsilon_{11}-1)-\varepsilon_{22}(\varepsilon_{11}+1)+(\varepsilon_{11}-1)(\varepsilon_{22}-1)\cos\left(2(\theta_1+\theta_2)\right)\right)=0
\end{aligned} \tag{S6}$$

It is straightforward to check that Eq. (S6) is equivalent to a sixth-degree polynomial in the variable $k_z$.

## B. LOSS REDISTRIBUTION

Figure S1 complements Fig. 2 of the main text by showing how losses are redistributed as the angle $\theta_2$ is varied. As in hyperbolic shear metasurfaces [S4-S5], changing the angle between the two wire arrays induces a redistribution of losses. Here, however, the redistribution has a distinct character: it does not simply correspond to increasing the attenuation of one arm while decreasing that of the opposite arm, as in hyperbolic shear metasurfaces [S4-S5]. Because opposite arms of each hyperbola are associated with different wire arrays, their arm-to-arm attenuation contrast is mainly set by the loss contrast between the arrays and remains approximately preserved, particularly away from the hyperbola vertices. As $\theta_2$ is reduced from $90^\circ$, both arms of H1 become more attenuated, whereas both arms of H2 become less attenuated except in the immediate vicinity of the vertices. Thus, varying $\theta_2$ redistributes loss between the two hyperbolic propagation channels, while approximately preserving the arm-to-arm attenuation contrast within each branch.

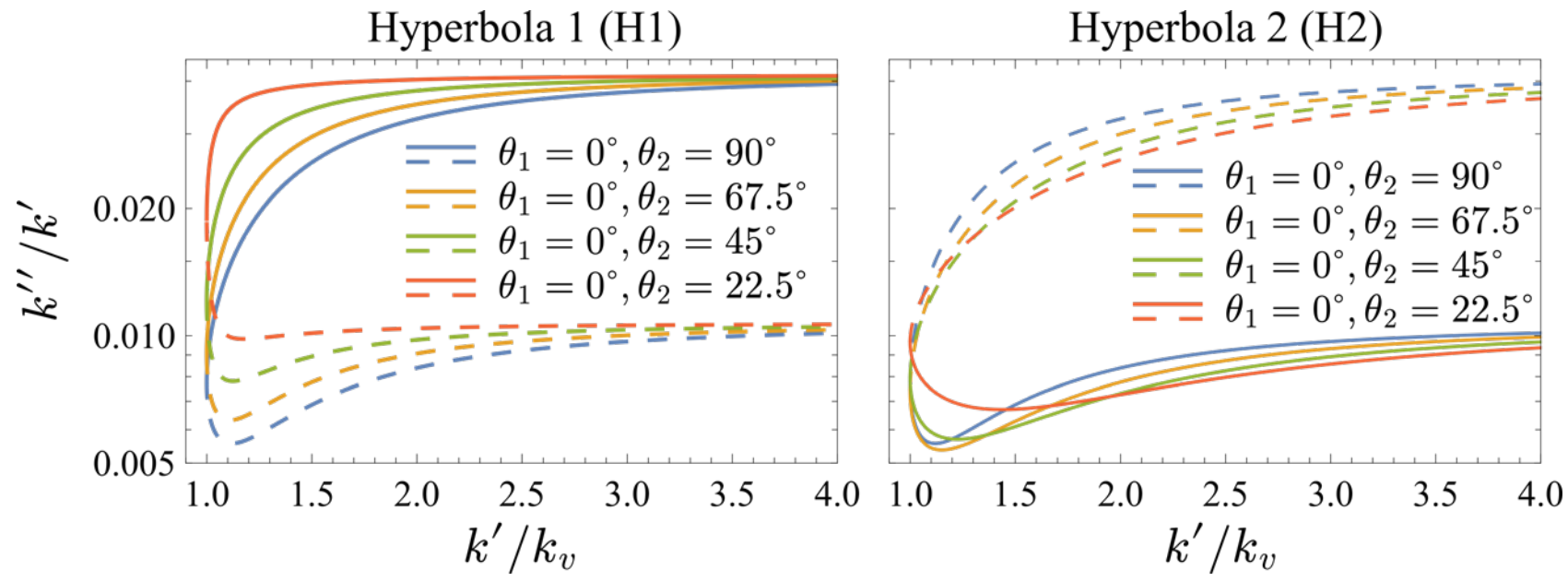

**Figure S1.** Branch-dependent loss redistribution along the hyperbolic contours as the angle $\theta_2$ is varied. Solid lines: right-hand arms of hyperbolas H1 and H2 (to the right of the vertices $k_v$); dashed lines: left-hand arms (to the left of the vertices $k_v$). H1 and H2 are defined in Fig. 2 of the main text. The lattice constant is $a = 2.5\ \mu$m and the wire radii are $r_{w,1} = r_{w,2} = 0.05a$. The wires of the arrays 1 and 2 follow a Drude model with plasma frequencies $\omega_{p,1}/(2\pi) = 2000$ THz and $\omega_{p,2} = 2\omega_{p,1}$, respectively, equal damping rates $\Gamma_1 = \Gamma_2 = 0.02\omega_{p,1}$, and high-frequency permittivities $\varepsilon_{\infty,1} = \varepsilon_{\infty,2} = 1$.

## C. SCATTERING PROBLEM

In this section, we study the plane wave scattering problem using the non-local homogenization model described in the main text of the article. The metamaterial slab is assumed infinite and periodic along the *x*- and *y*- directions, and finite along the *z*- direction (with thickness *L*). The incident wave propagates in the *xz* plane ($k_y = 0$) and the incoming magnetic field is polarized along the *y*-direction (see Fig. S2). Thus, the magnetic field in the three regions of space can be written as follows (the *x*-dependence and the time variation $e^{-i\omega t}$ are suppressed):

$$\begin{aligned} H_y^{(1)} &= H_{\text{inc}}(e^{\gamma_0 z} + R e^{-\gamma_0 z}), \quad z > 0 \\ H_y^{(2)} &= A_1 e^{ik_z^{(1)} z} + A_2 e^{ik_z^{(2)} z} + A_3 e^{ik_z^{(3)} z} + A_4 e^{ik_z^{(4)} z} + A_5 e^{ik_z^{(5)} z} + A_6 e^{ik_z^{(6)} z}, \quad -L < z < 0 \\ H_y^{(3)} &= H_{\text{inc}} T e^{\gamma_0 (z+L)}, \quad z < -L \end{aligned} \tag{S7}$$

In the above, $H_{\text{inc}}$ is the incident field, $\gamma_0 = -ik_{z0}$ (with $k_{z0} = \sqrt{k_0^2 - k_x^2}$) is the free-space propagation constant, $k_x = k_0 \sin\theta_{\text{inc}}$ is the transverse wavevector component, $k_0 = \omega/c$ is the free-space wavenumber, $k_z^{(l)}$ (with $l = 1,...,6$) are the propagation constants and $A_l$ (with $l = 1,...,6$) the complex amplitudes of the plane waves excited inside the double-wire medium, and $R$ and $T$ are the reflection and transmission coefficients, respectively. The propagation constants $k_z^{(l)}$ are obtained by solving Eq. (S6). For each plane wave with magnetic field of the form $\mathbf{H} = H_0 e^{i\mathbf{k}.\mathbf{r}} \hat{\mathbf{u}}_y$, the corresponding electric field is given by

$$\mathbf{E} = \frac{H_0}{\omega \varepsilon_0} \overline{\overline{\varepsilon}}_{\text{eff}}^{-1} \cdot \left( \hat{\mathbf{u}}_y \times \mathbf{k} \right) e^{i\mathbf{k}.\mathbf{r}} . \quad \text{(S8)}$$

Using Eq. (S8), it is straightforward to write the electric field in all space, similar to Eq. (S7).

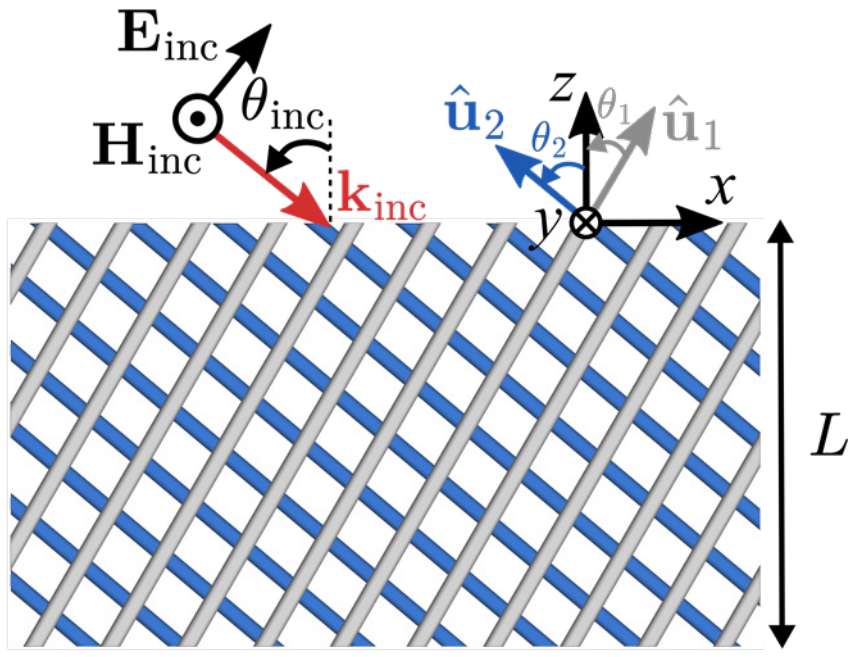

**Figure S2.** Setup for the scattering problem. A plane wave impinges on a heterogenous double-wire medium slab with thickness *L*. The incident wave vector $\mathbf{k}_{\text{inc}}$ and electric field $\mathbf{E}_{\text{inc}}$ lie in the *xz* plane, while the magnetic field $\mathbf{H}_{\text{inc}}$ is polarized along the *y*-direction.

To calculate the reflection and transmission coefficients, it is necessary to impose the following boundary conditions:

$$E_x \text{ and } H_y \text{ are continuous at } z = 0 \text{ and } z = -L , \quad \text{(S9)}$$

$$\mathbf{J}_{d,\text{av}}.\hat{\mathbf{u}}_1 = 0 , \ \mathbf{J}_{d,\text{av}}.\hat{\mathbf{u}}_2 = 0 \text{ at } z = -L^+ \text{ and } z = 0^- . \quad \text{(S10)}$$

The first set of boundary conditions corresponds to the classical ones and imposes the continuity of tangential electric and magnetic fields at the interfaces. The second set corresponds to the additional boundary conditions (ABCs) introduced in Refs. [S6-S7], which ensure that the electric current flowing along each metallic wire vanishes at both interfaces (see Ref. [S7] for the definition of the averaged current $\mathbf{J}_{d,\text{av}}$). These ABCs are necessary to handle the extra degrees of freedom characteristic of nonlocal (spatially dispersive) materials, manifested through the existence of “additional waves”. Thus, the scattering problem reduces to a linear system that can be readily solved numerically.

Solving this linear system yields the complex reflection and transmission amplitudes, $R$ and $T$, respectively. The corresponding power reflectance and transmittance are $\mathcal{R} = |R|^2$ and $\mathcal{T} = |T|^2$, while the absorptance follows from energy conservation as $\mathcal{A} = 1 - \mathcal{R} - \mathcal{T}$. These quantities are used in the main text to quantify the angularly asymmetric transmission and absorption.

## D. SYMMETRY OF THE SCATTERING PARAMETERS

Here, we analyze the symmetry relations of the scattering parameters of the heterogeneous double-wire medium slab. We consider different types of illumination of the metamaterial slab, as depicted in Fig. S3. The incoming wave may arrive either from the top or from the bottom air region, with incidence angle $\theta_{\text{inc}}$ or $-\theta_{\text{inc}}$, thereby defining four distinct propagation channels..

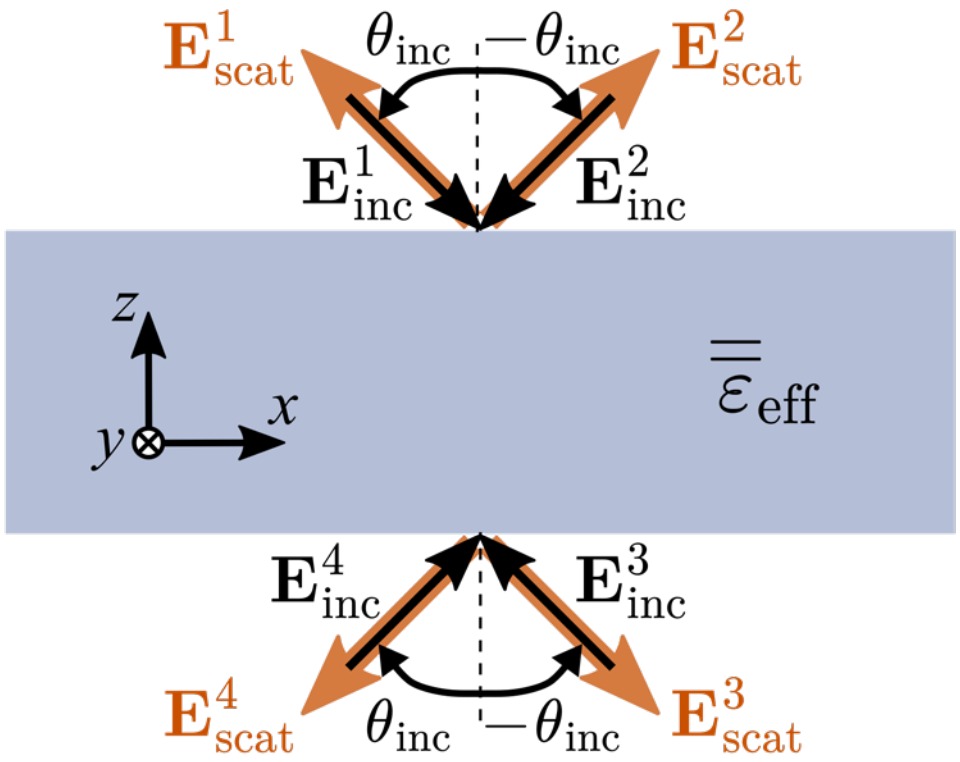

**Figure S3.** The metamaterial slab modeled as a four-port network. Each port corresponds to a specific propagation channel defined by the incidence angles $\pm\theta_{\text{inc}}$ on the top and bottom interfaces.

This system may be regarded as a four-port network, where each port is associated with a specific propagation channel. Since an incoming wave propagating in channel 1 generates only outgoing waves in channels 2 and 3, and so on, it becomes evident that the scattering matrix of the network has the form

$$\mathbf{S} = \begin{pmatrix} 0 & S_{12} & S_{13} & 0 \\ S_{21} & 0 & 0 & S_{24} \\ S_{31} & 0 & 0 & S_{34} \\ 0 & S_{42} & S_{43} & 0 \end{pmatrix}. \tag{S11}$$

Considering the $180^\circ$-rotation symmetry of the system about the *y*-axis, it is possible to link the scattering parameters for excitations of ports 1 and 3, as well as those for excitations of ports 2 and 4. This allows the scattering matrix to be written as

$$\mathbf{S} = \begin{pmatrix} 0 & R_2 & T_1 & 0 \\ R_1 & 0 & 0 & T_2 \\ T_1 & 0 & 0 & R_2 \\ 0 & T_2 & R_1 & 0 \end{pmatrix}. \tag{S12}$$

Here $R_1$ and $R_2$ denote reflection coefficients for incidence from $\theta_{\text{inc}}$ and $-\theta_{\text{inc}}$, respectively, for both top and bottom illumination. The coefficients $T_1$ and $T_2$ describe the transmission between the reciprocal channel pairs $1 \leftrightarrow 3$ and $2 \leftrightarrow 4$, respectively.

Because the system is fully reciprocal—it involves no external bias, motion, or nonlinear effects— the Lorentz reciprocity theorem requires the scattering matrix to be symmetric, i.e., $\mathbf{S} = \mathbf{S}^T$. Applying this condition yields that $R_1 = R_2$, which directly implies that the reflection coefficient is an even function of $\theta_{\text{inc}}$, i.e., $R(\theta_{\text{inc}}) = R(-\theta_{\text{inc}})$. This explains why the reflectance remains symmetric in the main-text figures, even though the medium exhibits strongly asymmetric dissipation.

By contrast, for the transmission coefficients, the reciprocity condition $\mathbf{S} = \mathbf{S}^T$ does not impose that $T(\theta_{\text{inc}}) = T(-\theta_{\text{inc}})$. Instead, it only enforces that the transmission from the top to the bottom of the slab ($T_{\text{t}\to\text{b}}$) equals the transmission from the bottom to the top ($T_{\text{b}\to\text{t}}$) under reversed illumination, i.e., $T_{\text{t}\to\text{b}}(\pm\theta_{\text{inc}}) = T_{\text{b}\to\text{t}}(\mp\theta_{\text{inc}})$. Therefore, for illumination from a fixed side, one may have $T(\theta_{\text{inc}}) \neq T(-\theta_{\text{inc}})$, without violating reciprocity.

The angular asymmetry in transmission and absorption observed in the main text therefore does not originate from reciprocity breaking. Rather, it arises because waves incident from opposite angles couple to different loss-asymmetric hyperbolic propagation channels of the heterogeneous double-wire medium. These channels have markedly different attenuation, leading to asymmetric transmission and absorptance for one-sided illumination, while reciprocity enforces symmetric reflection.

## E. GAUSSIAN BEAM EXCITATION

This section discusses the excitation of the heterogeneous double-wire metamaterial slab by a Gaussian beam with transverse magnetic polarization (TM-*z* polarized). We use the nonlocal homogenization model to characterize the scattering of the Gaussian beam at the interfaces of a metamaterial slab infinitely extended along the *x*- and *y*- directions.

We consider a cylindrical Gaussian beam with beam waist $w_0$, angle of incidence $\theta_{\rm inc}$, and focal point at $z = z_0$ in front of the metamaterial slab. The normalized magnetic field of the Gaussian beam excitation is represented as a continuous superposition of plane waves, i.e., through a spatial Fourier integral over its plane-wave spectrum,

$$H_y^{\rm GB}(x,z) = \int_{-\infty}^{\infty} H_y^{\rm GB}(\omega,k_x) e^{\gamma_0(z-z_0)} e^{ik_x x} dk_x \text{, where } H_y^{\rm GB}(\omega,k_x) = \frac{e^{-\frac{1}{4}\left(k_x - \frac{\omega}{c}\sin\theta_{\rm inc}\right) w_0^2} w_0}{2\sqrt{\pi}} \quad \text{(S13)}$$

The response of the system to this excitation is obtained by superposing the scattering contributions of each plane-wave component in the decomposition (S13). Thus, the magnetic field in the three regions of space can be written as follows:

$$\begin{aligned}
H_y^{(1)}(x,z) &= \int_{-\infty}^{\infty} H_y^{\rm GB}(\omega,k_x)(e^{\gamma_0(z-z_0)} + R(\omega,k_x)e^{-\gamma_0(z+z_0)})e^{-jk_x x} dk_x, \quad z > 0 \\
H_y^{(2)}(x,z) &= \int_{-\infty}^{\infty} H_y^{\rm GB}(\omega,k_x) H_y^{(2)}(k_x,z) e^{-\gamma_0 z_0} e^{-jk_x x} dk_x dk_x, \quad -L < z < 0 \quad , \\
H_y^{(3)}(x,z) &= \int_{-\infty}^{\infty} H_y^{\rm GB}(\omega,k_x) T(\omega,k_x) e^{\gamma_0(z+L-z_0)} e^{-jk_x x} dk_x dk_x, \quad z < -L
\end{aligned} \quad \text{(S14)}$$

where $H_y^{\mathrm{GB}}(\omega,k_x)$ is given by Eq. (S13), $H_y^{(2)}(k_x,z)$ is the magnetic field inside the slab ($-L<z<0$) and defined in Eq. (S7), and $R(\omega,k_x)$ and $T(\omega,k_x)$ are the reflection and transmission coefficients obtained by solving the plane wave scattering problem outlined in Sec. C. Using the above equations, we can calculate the magnetic field profile in all regions of space.

Figure S4 shows additional infrared results for a Ag double-wire medium at $\omega/(2\pi)=40$ THz. The wire radii and lattice constant are the same as in Fig. 4 of the main text, while $\theta_1=\theta_2=67.5^\circ$ and $L=15a$. As in the main text, the power coefficients exhibit a pronounced angular asymmetry: waves with $\theta_{\mathrm{inc}}<0$ are predominantly absorbed, with very small transmittance, whereas waves with $\theta_{\mathrm{inc}}>0$ are transmitted more efficiently.

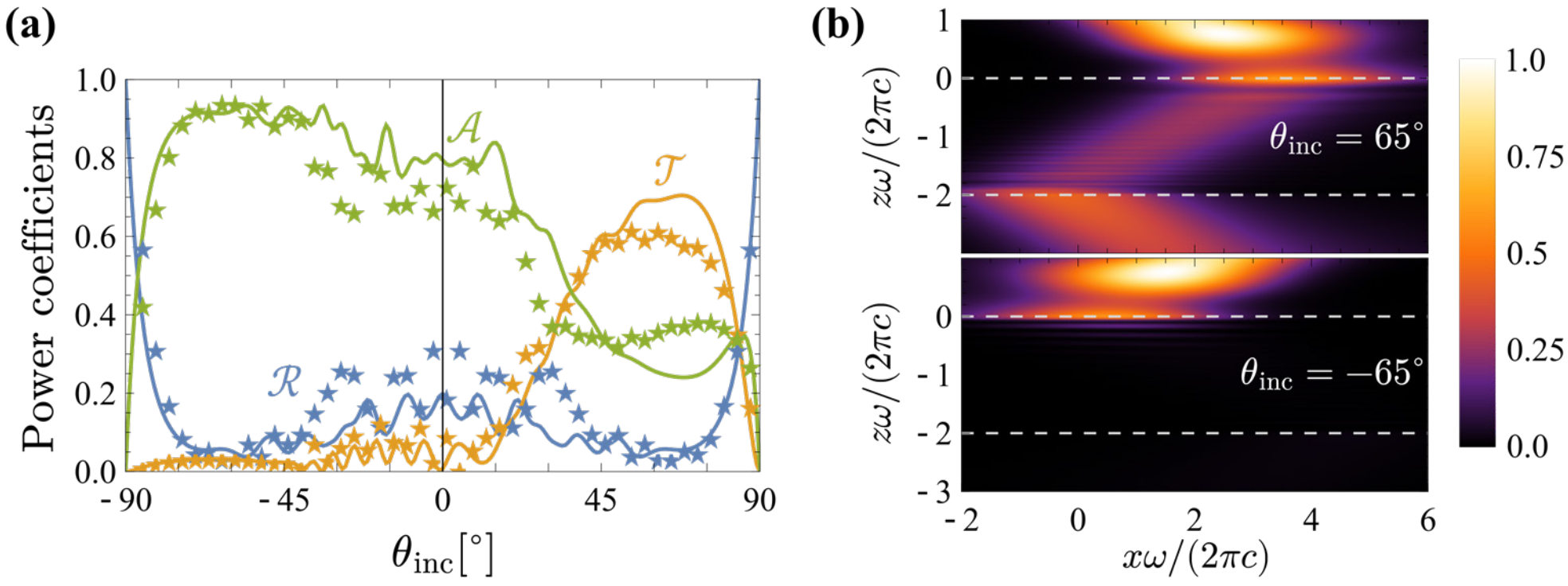

**Figure S4.** Similar to panels (b) and (c) of Fig. 4 of the main text, but for a Ag double-wire medium at $\omega/(2\pi)=40$ THz, $\theta_1=\theta_2=67.5^\circ$, and $L=15a$.

The corresponding Gaussian-beam simulations further confirm asymmetric negative refraction at this higher infrared frequency. For $\theta_{\mathrm{inc}}=65^\circ$, the beam is negatively refracted at both interfaces and transmitted through the slab. Conversely, for $\theta_{\mathrm{inc}}=-65^\circ$, the beam couples to the highly dissipative channel (associated with array 2) and is strongly attenuated inside the metamaterial. These results demonstrate that the angularly asymmetric transmission and absorption and the associated asymmetric negative-refraction persist even at higher infrared frequencies.

## F. ASYMMETRIC NEGATIVE REFRACTION IN FEW-LAYER DOUBLE-WIRE SLABS

To validate the asymmetric negative-refraction effects predicted by the nonlocal homogenization theory in the main text, we performed full-wave simulations of a heterogeneous double-wire metamaterial slab under Gaussian-beam illumination using CST Studio Suite [S8]. In contrast to the homogenized model, which assumes an infinitely extended slab, we consider here structures that are finite in both the *x*- and *y*-directions. Specifically, the slabs have a finite lateral width along *x* of $L_x = 90a$ (with $a = 2.5$ mm) and comprise only a few wire planes along the *y*-direction. We analyze slabs with four (Fig. S5(a)), six (Fig. S5(b)), and eight (Fig. S5(c)) wire planes, while all remaining parameters are identical to those used in Fig. 6 of the End Matter of the article.

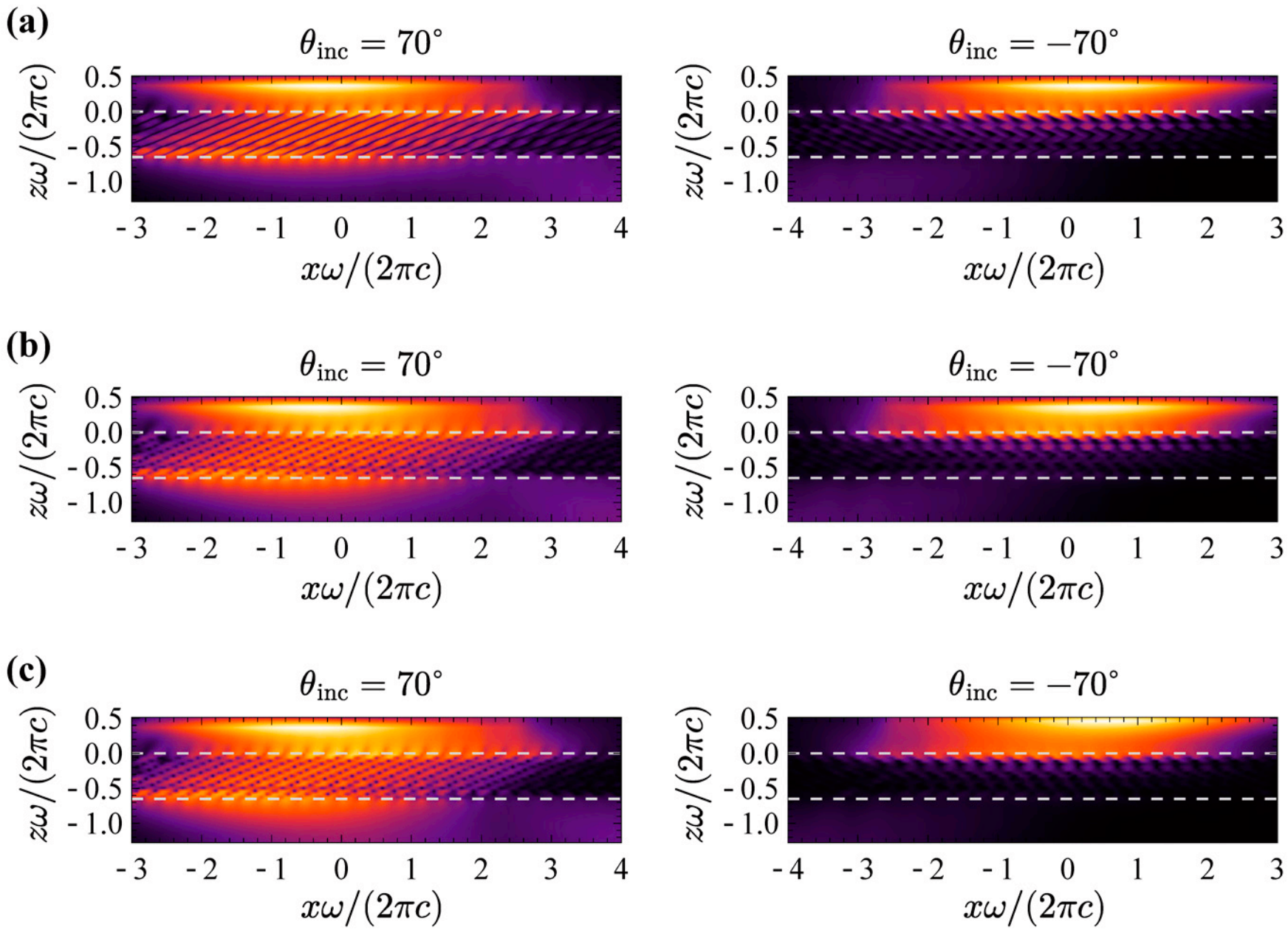

**Figure S5.** Normalized amplitude of the magnetic field ($|\mathbf{H}|$), obtained from full-wave simulations in CST Studio Suite [S8], for finite-width heterogeneous double-wire slabs comprising (a) four, (b) six, and (c) eight wire planes along the *y*-direction. The Gaussian-beam width along *y* is $1.8a$, $2.8a$, and $3.8a$ in panels (a), (b), and (c), respectively. The slabs have finite lateral width $L_x = 90a$. All other parameters match those of Fig. 6(c) of the End Matter of the article: $\omega/(2\pi) = 13$ GHz, $\theta_1 = \theta_2 = 67.5°$, $L = 6a$, and $w_0 = 4\pi c/\omega$. The dashed lines denote the slab interfaces.

Figure S5 shows the normalized amplitude of the magnetic field for Gaussian beams incident at (*i*) $\theta_{\text{inc}} = 70^\circ$ and (*ii*) $\theta_{\text{inc}} = -70^\circ$. These full-wave results are qualitatively consistent with the homogenization predictions of Fig. 6(c) in the End Matter of the article. In all cases, the finite slabs exhibit strongly asymmetric negative refraction: for positive incidence angles, the beam is negatively refracted and transmitted through the slab, whereas for negative incidence angles it is strongly attenuated within the structure.

Crucially, these simulations demonstrate that the asymmetric negative refraction persists even in structures comprising only a few wire planes, indicating that the effect is not limited to bulk-like implementations of the double-wire geometry. Increasing the number of wire planes from four to six and eight further improves the transmitted beam for $\theta_{\text{inc}} > 0$, while preserving strong attenuation for $\theta_{\text{inc}} < 0$. Overall, these results show that the asymmetric response is robust in finite, few-layer structures and can be realized in compact, practically implementable double-wire slabs.